\documentstyle[12pt]{article}

\def\eop{\vspace*{\fill}\pagebreak}
\begin{document}
 \begin{titlepage}
{\bf December 1992, }\hfill       {\bf PUPT-1369}\\
\begin{center}
 {\bf  THE THEORY OF TURBULENCE IN TWO DIMENSIONS }\\
 \vspace{1.5cm}
 {\bf  A.M.~Polyakov}\\
 \vspace{1.0cm}
 {\it  Physics Department, Princeton University,\\
Jadwin Hall, Princeton, NJ 08544-1000.\\
E-mail: polyakov@puhep1.princeton.edu}
 \vspace{1.9cm}
\end{center}
 \abstract{The methods of conformal field theory are used to obtain the series
of exact solutions of the fundamental equations of the theory of turbulence.
The basic conjecture, proved to be self-consistent ,is the conformal invariance
of the inertial range. The resulting physical picture is different from the
standard one , since the  enstrophy transfer is catalyzed by the large scale
motions. The theory gives some unambiguous predictions for the correlations in
the inertial range.}
\vfill
\end{titlepage}
 \section{Introduction}
  During the last thirty years we witnessed an amazing unification of physical
 ideas. Concepts and notions of the totally different regions of physics
appeared
 to be almost isomorphic. Spontaneous symmetry breaking and renormalization
 group are the most notable examples.\par
In this paper I extend slightly this set of isomorphic ideas. Namely, I apply
 methods of conformal field theory ( CFT ) to the problem of two-dimensional
 turbulence. Short version of this paper has been published earlier [1].\par
The major puzzle in the theory of turbulence is the following.
It is commonly believed, that at large Reynolds numbers we are dealing with
 the stationary statistical regime, describing velocity distribution. In other
 words, there should exist a time-independent probability
$ P=P[~v_{\alpha}\left( x \right)~]$
 (where $ v_{\alpha}\left( x \right)$      ) is the velocity). This probability
should commute with the
 equations of motion, or:
 \begin{equation}
\int{dx\left( {\delta P \over \delta v_{\alpha}\left( x \right)}
\right)\dot{v}_{\alpha}\left( x \right)}=0
\end{equation}
 (where it is understood that we express $ \dot{v}_{\alpha}$ through $
v_{\alpha}$    by the use of Navier - Stokes equations).
   At large Reynolds numbers viscosity can be neglected and hydrodynamics
becomes a hamiltonian theory with the hamiltonian:
 \begin{equation}\label{p}
H~=~\int{dx {1 \over 2}v_{\alpha}^{2}}
\end{equation}
 If so, the equation (1) means that  P    is the integral of motion for this
system.
 Since it is believed that in general there are no non-trivial integrals for
the system (2) (in two-dimensions there are, but these are irrelevant for our
discussion) we could erroneously conclude that:
 \begin{equation}\label{3}
P~=P\left( H \right)~=~exp\left\{ -\beta H \right\}
\end{equation}
  (the last equality follows from the usual additivity assumption). This is the
Gibbs distribution for the temperature $ \beta ^{-1}$     and it clearly does
not describe the  turbulent flow, in which we expect permanent energy flux.\par
  The resolution of this puzzle is that in fact there are extra integrals of
motion  which form $ P[~v_{\alpha}]$.      They are, however, highly non-local
and non-polynomial. It would be difficult and unnecessary to write them down
explicitly, because  there still be a problem of         averaging with respect
to very complicated distributions.
  In what follows we will mostly use the analogue of the equation (1) applied
directly to the correlation functions. Nevertheless  we  will briefly comment
on the origin of the extra integral of motions for the systems with weak
coupling. In this case they are just naive field-theoretic constants of motion
formed of " in" and "out" operators.
  When expressed in terms of the Heisenberg fields they become non-local, and
describe the distributions of weak turbulence.\par
 The central point of this paper is, however, quite different. We will consider
the equations for the correlation functions:
 \begin{equation}\label{4}
<~\dot{v}_{\alpha_{1}}\left( x_{1} \right)v_{\alpha_{2}}\left( x_{2}
\right)\cdots~>+<~v_{\alpha_{1}}\left( x_{1} \right)\dot{v}_{\alpha_{2}}\left(
x_{2} \right)\cdots~>+\cdots=0
 \end{equation}
  (where   $ \dot{v}_{\alpha}$   is expressed in terms of   v   by the use of
equations of motion). These are standard equations (The Hopf equations [2] )
which express N - point functions
   in terms of the N + 1 - point function.
The existing procedures for dealing with these equations are based on some kind
of closure hypothesis expressing, say the 4 - point function as a square of 2 -
point function.\par
  We take a different approach. Namely, we try to satisfy (4) exactly, by the
use of conformal field theory, assuming that the developed turbulence in the
inertial range possesses infinite conformal symmetry.
  There is no proof that this must necessarily the case, and that our conformal
 solutions are the only possible ones. What we have done is only a successful
ansatz which solves exactly eqs (4).
  \section  { Chains of equations solved by conformal field theories.}
In this section I will sum up some basic properties of CFT, needed below  as a
 representative example, I will use the most familiar case of the Ising model.
 The Ising model is a  $  \varphi^{4}$ -field theory[3]. That means that if we
treat $ \varphi$ as a fluctuating variable, it satisfies the equation:
  \begin{equation}\label{5}
\partial^{2}\varphi\left( x \right)+\mu \varphi\left( x
\right)=g\varphi^{3}\left( x \right)
 \end{equation}
implying the relations between N - point functions of  $ \varphi$ and N + 3 -
point functions,
   known as Schwinger - Dyson equations. The constant  $ \mu_{0}$ must be fine
-
 tuned in order to put  the system to the critical point, or in order to allow
(5) to have conformal invariant solutions.
 Now let us try to answer the basic question of this section : how CFT solves
eqs (5)?
 In CFT we are dealing directly with the sets of correlation functions at the
critical point. The basic assumptions [4] are that there exists a set of the so
called
 primary operators $ \left\{ O_{k}\left( x \right)\right\} $
 with anomalous dimensions  $ \left( \Delta_{k}, \bar{\Delta} _{k}\right)$
 (referring to the rescaling of variables $ z ~=~x_{1}+i x_{2}$ and  $
\bar{z}~=~x_{1}-ix_{2}$.
 It is assumed that under arbitrary analytic transformation $ z
{}~\Rightarrow~f\left( z \right)$ ~,$ O_{k}$
 transform as:
 \begin{equation}\label{6}
O_{k}\left( z,\bar{z} \right)\Longrightarrow\left( {\partial {f}\over \partial
z} \right)^{\Delta_{k}}\overline{\left({\partial {f}\over \partial
z}\right)^{\Delta_{k}}}~O_{{k}}\left( f,\bar{f} \right)
  \end{equation}
 This transformation is generated by the analytic energy-momentum tensors T(z)
and $ \bar{T}(z)$
 (related to the tensor $  T _{\alpha\beta}\left( x \right) $ as $
T,\bar{T}~=~T_{11}-T_{22}\pm 2iT_ {12})$.
  Conformal invariance implies that the trace of the energy-momentum tensor is
zero.\par
 The basic result of [4]  was that together with primary operators, CFT is
 bound to contain the so-called secondary or descendent operators. They come
 from the conformal deformations of the primary ones.
 In order to describe them it is convenient to introduce the Virasoro algebra:
\begin{equation}\label{7}
L_{n}= \oint{~dz~z^{n+1}T\left( z \right)
}
 \end{equation}
 \begin{equation}\label{8}
[~L_{n}, L_{m}~]~=~\left( n-m \right)L_{n+m}+{c \over 12}n\left( n^{2}
-1\right)\delta_{n+m,0}
\end{equation}
 The secondary operators are given by:
$ O_{k}^{n_{1}\cdots n_{l},m_{1}\cdots m_{n}}~=L_{-n_{1}}\ldots
L_{-n_{l}}\overline{L_{-m_{1}}\ldots L_{-m_{n}}}O_{k}$
 The correlation functions of the secondaries are expressed by the differential
 relations through the correlators of the primaries [4].\par
 The second basic feature of CFT is the operator product expansion. It says
that
 \begin{equation}\label{}
O_{k}\left( z \right)O_{l}\left( 0 \right)~=~\sum{f_{kl}^{m}\left( z\bar{z}
\right)^{\Delta_{m}-\Delta_{k}-\Delta_{l}}O_{m}\left( 0
\right)+\mbox{secondaries}}
\end{equation}
 The structure constants
 satisfy simple (in principal) consistency relation coming from the
associativity
 of the algebra. Thus, the classification problem of CFT is reminiscent of the
 classification problem of the Lie algebras, but is vastly more complicated.
\par
 However, in [4] we have found a simplest set of CFT's, the so called minimal
 models. In these models the number of the primaries is finite (but the total
 number of operators is, of course, infinite).
 The Ising model is the minimal model of the type (3,4) with only 3 primary
 operators. They are unit operator I ,    spin $ \sigma$ and energy density
  $\varepsilon$.
 The fusion rules (9) in this case are:
 \begin{eqnarray}
[\sigma]~[\sigma]~=~[I]+[\varepsilon]\\~[\sigma]~[\varepsilon]~=[\sigma]\\~[\varepsilon][\varepsilon]~=~[I]
\end{eqnarray}
 The brackets in these symbolic formulas mean the "conformal class", namely
 the primary operator itself together with all it descendents.\par
 Now we are coming to the central point. Let us identify the field~$ \varphi$
 of the $ \varphi ^{4}$~theory with the field $ \sigma$ of the conformal field
theory.
 We have to examine the quantities \\
$ <\varphi^{3 }\left( x \right)~\varphi\left( y_{1} \right )
\cdots\varphi\left( y _{n}\right ) >$.
 However, the conformal field theory is applicable only when all points are
well
 separated compared to the lattice spacing. The way out of this problem is to
 introduce the point - splitted definition of the $ \varphi^{3}\left( x
\right)$.
 Let us examine the object:
 \begin{equation}\label{}
\overline{\lim_{a\rightarrow 0}{}}\sigma\left( x+{a \over 2}
\right)\sigma\left( x-{a \over 2} \right)\sigma\left( x
\right)\equiv"\sigma^{3}\left( x \right)"\stackrel{\rm def}{=}\Phi\left( x,a
\right)
\end{equation}
 The symbol $\overline{lim} $  here means that we first of all average over
directions of $a$
 and then take  $|a| $ to be much smaller then all other distances $ |x-y_{i}|$
 and  $| y_{i}-y_{j}|$.
 Keeping it at first much larger than
 the lattice spacing, we can apply the
 fusion rules (9).  Since
 \begin{equation}\label{}
[\sigma]~[\sigma]~[\sigma]~=~[\sigma]
\end{equation}
  We obtain:
 \begin{equation}\label{}
\Phi\left( x,a \right)\approx|a|^{-4\Delta_{\sigma}}[c_{1}\sigma\left( x
\right)+c_{2}a^{2}\partial^{2}\sigma\left( x \right)+\ldots]
\end{equation}
  Although this equation is correct only for $ |a|~>>~l$
 (where  l is a lattice spacing), with any natural definition of the theory it
can
 be extrapolated to  $ |a|\sim~l$.
    The reason is that with any isotropic regularization at the distance
 we will get instead of (15):
 \begin{equation}\label{}
\Phi\left( x,a \right)=l^{-4\Delta_{\sigma}}f\left( {|a| \over l}
\right)\sigma\left( x \right)+\ldots
\end{equation}
 (where precise form of  f  depends on the way we regulate the theory).
   As a
 result, if we take $ a{\
\lower-1.2pt\vbox{\hbox{\rlap{$<$}\lower5pt\vbox{\hbox{$\sim$}}}}\ }l$ we still
have the equation (15), and hence the basic field equation (5) is satisfied.
  All this means one simple thing -field equations of motion, when read from
the
 right to the left are nothing but operator product expansions. This fact was
 well understood from the very beginning of the field theory for critical
phenomena.
    Point-splitting and directional averaging over $ a$  is also nothing new -
 already in quantum electro-dynamics it has been necessary to introduce the
 isotropic cut-off in the momentum space, which is equivalent to the procedure
 we discuss.\par
 To summarize - in this section we have seen how the structure of the operator
 product expansion reflects equations, satisfied by the correlation functions.
  \section{Solutions of the inviscid Hopf equations by the conformal field
theory.}
 Let us turn now to the Hopf equations (4) and try to solve it in the spirit of
the
 proceeding section. In two dimensions it is convenient to introduce vorticity
$ \omega$  and the stream function  $ \psi$ given by
\begin{eqnarray}\label{}
   v_{\alpha}\left( x \right)~=~e_{\alpha\beta}\  \partial_{\beta}\psi~~~
\omega\left( x
\right)~=~e_{\alpha\beta}\partial_{\alpha}v_{\beta}~=~\partial^{2}\psi
\end{eqnarray}
 They satisfy Navier - Stokes equations:
 \begin{equation}\label{}
\dot{\omega}~+~e_{\alpha\beta} \partial_{\alpha}\psi~
\partial_{\beta}\partial^{2}\psi~=~\nu \partial^{2}\omega
\end{equation}
 If the stirring force is present, it must be added to the RHS of (18).
 If we assume, that for the large Reynolds numbers there exist the inertial
range
 of scales in which both viscosity and the stirring force are negligible, we
can as
 a first step examine the inviscid Hopf equation:
 \begin{eqnarray}\label{}
  <\dot{\omega} \left( x _{1}\right) ~\omega\left( x_{2}
\right)\cdots>~+~<\omega\left( x_{1} \right)~\dot{\omega}\left( x_{2}
\right)~\cdots>~+~\cdots~=~0\\
  \dot{\omega}\left( x \right)~=~-~"e_{\alpha\beta}\partial_{\alpha
}\psi\partial_{\beta}\partial^{2}\psi"
 \end{eqnarray}
  Here, just as in the previous section, we must be careful in defining the
correlators at the coinciding points. The reason there have been tied to the
existence of
  the lattice cut-off (where conformal theory fails).
 The reason here is analogous - we expect that in the momentum space only $
\omega\left(  {k } \right)$
 with
$ {1 \over L}~<~| {k}|~<{1 \over a}$ should satisfy (19). Here a  is the
ultraviolet cut-off proportional, as we see later, to some power of the
viscosity $ \nu$  while L  is the infrared cut-off, dictated  by the scale of
the stirring force.
  The UV - cut - off in the momentum space means that we must use the point-
 splitted definition of the product in (20)
\begin{equation}
"e_{\alpha\beta}\partial_{\alpha}\psi\left( x
\right)\partial_{\beta}\partial^{2}\psi\left( x
\right)"=\overline{\lim_{a\rightarrow
0}{}}e_{\alpha\beta}\partial_{\alpha}\psi\left( x+a
\right)\partial_{\beta}\partial^{2}\psi\left( x \right)
\end{equation}
 Let us assume now, that $ \psi$ is a primary operator of some yet unknown
conformal
   field theory and compute the RHS of (21) by the use of operator algebra.
 Suppose that we have the structure:
 \begin{equation}\label{}
\psi\left( x+a \right)\psi\left( x \right)~=~\left( a\bar{a}
\right)^{\Delta_{\phi}-2\Delta_{\psi}}\left\{ \phi\left( x
\right)+\mbox{descendents} \right\}
\end{equation}
  Here we introduced the operator  $ \phi$ which is the minimal dimension
operator
 in the operator product (22).
 In unitary theories (as in the Ising model) it is a unit operator.
 However, only Gibbs states (from which "nothing disappears") are described
 by the unitary theories. In turbulence we have "flux states" from which
 conserved quantities leak away, and it is natural to expect that they will be
 described by the non-unitary theories. In this case most of operators have
negative
   dimensions and  $ \phi$  is non-trivial.
 It is clear that when the operation (21) is applied to (22), the leading term
in
 the expansion gives zero.
 This happens because the result must be pseudoscalar and we can't form it out
 of  $ \phi$  and its derivatives. Hence we have to consider  subleading
descendents
 in (22). They have the general structure:
 \begin{eqnarray}
\mbox{descendents}=\sum{c_{\left\{ n \right\}\left\{ m
\right\}}}L_{-n_{1}}\ldots L_{-n_{k}}\overline{L_{-m_{1}}\ldots
L_{-m_{l}}}\phi\left( x
\right)\nonumber\\({a\bar{a}})^{\Delta_{\phi}-2\Delta_{\psi}}{a}^{\sum{n}}{\bar{a}}^{\sum{m}}
\end{eqnarray}
  Differentiation and directional averaging in (21) the leading term in (23):
  \begin{equation}\label{}
"e_{\alpha\beta}\partial_{\alpha}\psi\left( x
\right)\partial_{\beta}\partial^{2}\psi\left( x
\right)"=\mbox{const}({a\bar{a}})^{\Delta_{\phi}-2\Delta_{\psi}}~[L_{-2}\bar{L}_
{-1}^{2}-\bar{L}_{-2}L^{2}_{-1}]~\phi
\end{equation}
 This one of the basic formulas of this work.
 The reason, why this descendent appeared in (24) is quite simple. The LHS of
 (24) is pseudoscalar and hence it changes sign under complex conjugation.
 The lowest descendent which has the same property is the one we find in (24).
 In deriving (24) we accounted for the UV - divergency, but disregarded the IR
 one. We will return to the infrared side of the problem below.\par  Let us
assume for
 the moment that IR - divergencies are not present.
 If so, we derive a rather amazing conclusion from the eq.(24).
 Namely, there are two possibilities. The first one is $
\Delta_{\phi}~\leq~2\Delta_{\psi}$
 (negative:defect of dimensions").
 In this case the only way to satisfy (19) would be to assume that the operator
 \begin{equation}\label{}
\Omega~=~\left( L_{2}\bar{L}^{2}_{-1}~-~\bar{L}_{-2} L^{2}_{-1}\right)~\phi
\end{equation}
  is either zero or is a symmetry of the underlying CFT.
 The latter statement just stresses the fact that the relation (19) implies
that the
 infinitesimal variation of $ \omega$  ,  $ \dot{\omega}~=~\Omega$      must
not change the correlation functions,
   or as in the  case of decaying turbulence rescale them.
 It is quite possible that non-trivial examples of such symmetries exist.
 However if we restrict ourselves with the simple minimal models, we will have
 to conclude that the RHS of (25) must be zero.
 This is easy to achieve if we require that the minimal dimension operator $
\phi$
 is degenerate on the level  two.
 That means that it satisfies the equation [4]:
 \begin{equation}\label{}
 \left( L_{-2} ~-~{3 \over 2\left( 2\Delta_{\phi} +1\right)}L^{2}_{-1}\right)
{}~\phi~=~0
\end{equation}
 If so,  $ \Omega$ = 0 . The simplest example, in which this relation is
satisfied is the
 minimal model (2 , 5) with $ \Delta_{\phi}=\Delta_{\psi}=-{1 \over 5}$\par
 Even more surprising is the fact, that if $ \Delta_ {\phi}~>~2\Delta_{\psi}$
 then no extra restrictions are needed because as  $ a~\rightarrow~0$ the RHS
of (25) is
 zero, and the inviscid Hoft equation is satisfied.
 So, any CFT with the positive defect of dimensions solves the Hopf
equations.\par
 This is a "turbulent" counter part of the fact that there are continuosly many
 parasite solutions of the static Euler equations.
 Let us remember that if we look at the "laminary" solutions of
 \begin{equation}\label{}
e_{\alpha\beta}\partial_{\alpha}\psi\partial_{\beta}\partial^{2 }\psi~=~0
\end{equation}
 we find that if
 $ \partial^{2}\psi~=~f\left( \psi \right)$
 with arbitrary   $ f\left( \psi \right)$ then (27) is satisfied. It is well
known, however,that matching
 with the viscous region by the boundary layer consideration removes this
 ambiguity.
This analogy makes it clear that the crucial point for selecting correct
solutions
 of the inviscid Hopf equations must be matching relations at the UV region of
 the cut-off momenta, where viscosity takes over. It is also clear, that since
we
 are dealing with negative dimensions, we have to analyze the infrared diver
 gencies.We can say that in turbulence there are two boundary layers in the
 "momentum space", each requiring a separate theory.
 In the following sections we will discuss some partial results in this
direction.
 \section{The infrared problem }
  The CFT's which are most interesting for the turbulence problem contain
negative dimensions. That means that the formal correlation functions are
positive
 powers of the distance. It is clear that we have to be very careful in the
infrared
 region and treat it separately.
Let us begin with the 2 - point function $ <\psi\left( 0 \right)\psi\left( x
\right)>$.
 When the dimension  $ \Delta_{\psi}<0$, CFT gives
 \begin{equation}\label{}
<\psi\left( 0 \right)\psi\left( x \right) \stackrel{\rm\left( conf
\right)}{>}~=~-|x|^{4|\Delta_{\psi}|}
\end{equation}
 This is clearly unphysical. The correct prescription should take into account
 the fact, that only momenta in the inertial range are described by CFT. So, we
 should expect that, if we take the Fourier transform of (28)
 \begin{equation}\label{}
<\psi\left( k \right)~\psi\left( -k \right)>~=~const{1 \over
|k|^{2+4|\Delta_{\psi}|}}
\end{equation}
 this result must be used only if $ L^{-1}<|k|<a^{-1}$.
 We do not know the contribution from the $ |k|\sim L^{-1}$.
 The only thing which can be said is that when returning to the  x  space it
   should give expressions, analytic in  $ x^{2}$.    So, we have:
 \begin{equation}\label{}
<~\psi\left( 0 \right)\psi\left( x \right)~>\sim\int_{|k|{\
\lower-1.2pt\vbox{\hbox{\rlap{$>$}\lower5pt\vbox{\hbox{$\sim$}}}}\
}L^{-1}}{e^{ikx}|k|^{-2-4|\Delta_{\psi}|}}\sim\left( c_{1}L^{4|\Delta_{\psi}|}
+\ldots \right)-|x|^{4|\Delta_{\psi}|}
\end{equation}
 with the first bracket representing the contribution of the infrared modes.
When we turn to the multi - point functions, the situation is essentially the
 same. When we examine their momentum space form:
 \begin{equation}\label{}
G\left( k_{1}\cdots k_{N} \right)~=~<\psi\left( k_{1} \right) \cdots\psi\left(
k_{N} \right)>~~~
\left( \Sigma k_{j}~=~0 \right)
\end{equation}
 we will assume that the CFT formulas for this quantity work,provided that:
\begin{equation}\label{}
L^{-1}<|k_{i}|<a^{-1}
\end{equation}
 and
 \begin{equation}\label{}
L^{-1}<|k_{i 1} +\cdots+k_{i l}|<a^{-1}
\end{equation}
  The (33) condition means that we have no small momenta transferred in any
 channel.
 Again, in the coordinate space all that means that the physical N - point
function
  is equal to the conformal one plus the terms, analytic in some
$ \left( x_{i}~-~x_{j} \right)$.
  Notice, that if we understand the momentum integrals as analytically
continued
  in  $ \left( \Delta_{\psi} \right)$ from the region, where they converge, we
get precisely conformal
 answer in the  x -  space.
 In order to get the physical correlator within this analytic regularization,
one
 has to add  $ \delta$ - functions in the  k - space, representing the infrared
modes.
 For instance, within the analytic prescription, we have instead of
(29)
 \begin{eqnarray}\label{}
<\psi\left( k \right)\psi\left( -k \right)>~=~{const \over
|k|^{2+4|\Delta_{\psi}|}}~+~ a_{1 }L^{4|\Delta_{\psi}|}\delta\left( k
\right)+a_{2}L^{4|\Delta_{\psi} |-2}  {\partial^{2} \over \partial k^{2}}
\delta\left(k  \right)~+~\cdots
\end{eqnarray}
  When Fourier - transformed, (34) gives again (30). \par
 So, the convenient way to summarize this situation is to say, that we have
some
 sort of Bose - condensate in the momentum space, formed by the large scale
 motions. The physical correlators differ from the conformal ones by the
condensate
   terms (we will call them $  \delta $- terms later).
 The $  \delta$ - terms occur whenever any sum of momenta in the correlator is
equal
 to zero. In the coordinate space they are represented by the terms analytic in
$ \left( x_{i}-x_{j} \right)$.
 Occurrence of the   $  \delta $- terms is not surprising , because we have
strirring forces acting at k=0.
 The main problem with them is that we do not know yet how to determine
 their form from the dynamics.
 The other related problem, which we will consider now, is whether they
 destroy our solution of the Hopf equations. The danger comes from the fact
 that in the sect.3 we dealt with the purely conformal correlators, and this
deri
 vation has to be reconsidered.
 In order to do that, let us rewrite our basic equation in the momentum space:
 \begin{equation}\label{}
<\dot{\omega}\left( {q} \right)\omega\left(  {f} _{1}\right)\cdots\omega\left(
{f}_{n} \right)>~=~
-\int{d^{2}k[k , q]\left( \left( q-k \right)^{2} -{k}^{2}\right)}<~\psi \left(
k \right)\psi \left( q-k \right)\omega \left( f_{1} \right)\cdots\omega \left(
f_{n} \right)~>
\end{equation}
  Generally speaking, this integral has infrared divergency as $ k\rightarrow0$
(or $ k\rightarrow q $ ).
 The results of the proceeding section guarantee that if we treat this IR
divergency
   by the analytic regularization, the RHS of (35) will be zero, as before.\par
 The problem now is to study what happens when we simply impose the IR cut  off
in (35). In order to do this, we need to know the behavior of the correlations
functions when one of the momenta is much smaller then the others.
 Let us first evaluate the divergencies at  $ k\rightarrow 0$.
  This limit in the
coordinate space
 is dominated by the large R limit in the Fourier transform:
 \begin{equation}\label{}
<\psi\left( k \right)~\psi\left( q-k \right)\omega\left( f_{1}
\right)\cdots\omega\left( f_{N} \right)>~=~ \int{d^{2}Re^{ikR}}~<\psi\left( R
\right)\psi\left( 0 \right)\omega\left( f_{1} \right)\cdots\omega\left( f_{N}
\right)>
\end{equation}
 The large R behavior is governed by the fusion rule, where we replace all
operators except $ \psi\left( R \right)$ by  $ \psi\left( 0 \right)$ .\   Using
notations  of  [4], we have the formula:
 \begin{equation}
<\psi\left( R \right)\psi\left( 0 \right)\omega\left( f_{1} \right)\ldots
>~=~\sum_{N}{<0|\psi\left( R \right)|\Delta+N><\Delta +N|\psi\left( 0
\right)\omega\left( f_{1} \right)\ldots |0>}
\end{equation}
  where  the states $ |\Delta + N>$  correspond to the secondary operators
of
 the  $ \psi$. Using the fact that
  \begin{equation}\label{}
<0|\psi\left( R \right)|\Delta+N>\infty{1 \over R^{2\Delta+N}}
\end{equation}
  we obtain the following leading terms in the asymptotic expansion
 \begin{eqnarray}\label{}
   <\psi\left( k \right)\psi\left( q-k \right)\omega\left( f_{1}
\right)\cdots\omega\left( f_{N} \right)>\approx{c_{1}
\over|k|^{6+\lambda}}~<\Delta| \psi\left( q \right)\omega{f_{1}}
\cdots\omega\left( f_{N} \right)|0>\nonumber\\~+c_{2} \left\{ {k_{+} \over
|k|^{6+\lambda}} <\Delta|L_{1}\psi \left( q \right)\omega\left( f_{1}
\right)\cdots\omega\left( f_{N} \right)\ 0 > + c.c. \right\} +\cdots
 \end{eqnarray}
 When we substitute (39) into (35), the first leading term drops out after
integration  over directions of k the second term does contribute. Simple
computation  gives:
\begin{equation}\label{}
  <\dot{\omega}\left( q \right)~\omega\left( f_{1} \right)\cdots\omega\left(
f_{n} \right)>=const   L^{2+\lambda} ~  \left\{ q_{+}<\Delta|L_{1}\omega\left(
q \right)\cdots\omega\left( f_{n} \right)| >-c.c. \right\}~+~O\left(
L^{\lambda} \right)
  \end{equation}
  This leading term can be interpreted as fluctuating transport by the large
scale
 eddies. Indeed, the relation (40) is equivalent to the operator equation:
 \begin{equation}\label{}
\dot{\omega}\left( x
\right)=L^{2+\lambda}\hat{u}_{\alpha}\partial_{\alpha}\omega\left( x \right)
\end{equation}
  Here we introduced a new  $   x  $ - independent operators  $
\hat{u}_{\pm}$by the rule:
 \begin{equation}\label{}
<0|\hat{u}_{+}\omega\left( x_{1} \right)\ldots \omega\left( x_{n}
\right)|0>~=~i<\Delta_{\psi}|L_{1}\omega\left( x_{1} \right)\ldots \omega\left(
x_{n} \right)|0>
\end{equation}
  and$ \hat{u}_{-}=\hat{u}^{*}_{+}$.
 It is clear, that although this infrared counterterm makes  $ \dot{\omega}\neq
0 $, the Hopf
 equation is still satisfied, due to the translation invariance.
 As we go to the order of $ L^{\lambda}$counterterms we find more complicated
forms. By
 the same technic it is possible to show, that the complete divergencies give
the following equation:
 \begin{equation}\label{}
\dot{\omega}=\hat{u}_{\alpha}\partial_{\alpha}\omega~+~\hat{h}_{\alpha\beta}\partial_{\alpha}  \partial_{\beta}\psi
\end{equation}
  with the
 \begin{equation}\label{}
<0|\hat{h}_{++}\omega\left( x_{1} \right)\cdots\omega\left( x_{N} \right)|0>~
\sim ~<\Delta|\left( L_{2}+\lambda L^{2}_{1} \right)\omega\left( x_{1} \right)
\cdots\omega\left( x_{N} \right)|0>
 \end{equation}
  (where  $ \hat{h}_{\alpha\beta}$- is a traceless tensor).
 This contribution does not cancel in the Hopf equation. There are also other
 infrared counterterm with this property, occurring from the region in the (45)
where:
$| k+f_{i1}+ \cdots +f_{il} |\rightarrow 0 $
  (for some choice of $ \left\{ f_{i} \right\}$ ).
 All that means that if we had defined the physical correlation functions by
simply
 cutting of the momentum integrals in the Fourier transform, the Hopf
 equation would be destroyed.
 On the other hand, if we had defined all integrals by analytic regularization
it
 would mean the identification of the physical and conformal correlators.
 In this case the Hopf equations is satisfied, but the positivity is broken.
 The way out of this problem is to find the $ \delta$ - terms which do not
spoil the
 Hopf equation but restore the positivity.
 The most blatant violation of positivity is easily cured, if we postulate that
any correlation function contains the constant part, coming from the zero
modes:
  \begin{equation}\label{}
<O_{n1} \left( x_{1} \right)O_{n2}\left( x_{2} \right) \cdots O_{nN}\left(
x_{N} \right) >^{phys}~=~c_{n1} \cdots n_{N}\left( L^{-2\sum_  {j}\Delta_{nj}}
\right)~+~<O_{n1}\cdots O_{nN}>^{conf}
\end{equation}
  (where C - are some constants).
 This type of $ \delta$- term does not contribute to the Hopf equations at non
-
 zero momenta, and at the same time prevails over the negative conformal
contributions, since   $ |x_{i}~-~x_{j}|~\ll L$.
 That doesn't mean that there could be no other IR counterterms. In fact it is
 possible to show that there are many types of the  $ \delta$ - terms,
consistent with
 the Hopf equations at
$ q_{i}\neq0$.
   In order to determine them we need to
 include the stirring force terms (which
 arise at  $ q_{i}=0$) and to use the matching relations. This is the problem
for  the
 future work.
 At present it is not quite clear to what extent the $ \delta$- terms are
universal i.e.
 independent of the large scale structures.\par
 There is another related issue which we should discuss in this
section.\footnote{I am grateful to A. Zamolodchicov for drawing my attention to
it}
 It is the question of the vacuum expectation values of different operators.
 When defining our conformal contributions, we have implicitly assumed that
 all these expectation values (VEV) are zero. In general, however we have:
 \begin{equation}\label{}
<O_{n}\left( x \right)>~=~C_  {n}\cdot{L^{-2\Delta_{n}}}
\end{equation}
  If $ \Delta_{n}~<~0$ , and $c_{n} \neq 0$
  these VEV considerably modify correlation functions, as was noticed by Al.
 Zamolodchikov[5 ]. The reason for that is simple. From the OPE we get:
 \begin{equation}\label{}
<\psi\left( r \right)\psi\left( 0 \right)>~=~r^{-4\Delta\psi}  <I>~+~r^
{2\left( \Delta_{\phi} -2\Delta_{\psi}\right)}~<\phi>+ \cdots =
r^{4|\Delta_{\psi}|}~\left\{ 1+<\phi>\cdot\left( {L \over r}
\right)^{2|\Delta_{\phi}|} ~+~\cdots\right\}
\end{equation}
 The second term in (48) is the dominating one.
 The question, whether VEV are non-zero is determined by the infrared boundary
 conditions. Our assumption that $ c_{n}~=~0$
 is not fundamental. Actually our consideration of the Hopf equations remains
 intact even if $ c_{n}\neq 0$.
 The only thing which becomes more complicated is the infrared divergency.
 Also, the UV - matching of the next section requires modified
consideration.\par
 The question of the  $ c_{n}$ values belongs to the same category of the IR
problems
 which we left for the future work.
 At present the situation resembles quantum chromodynamics, where we have
 not solved the IR problem, but can, nevertheless, test its small distance
behaviour.\par
Let us summarize. The true physical correlators can be obtained from the
conformal
 ones either by defining the momentum space integrals in the sense of
 analytic regularization or in the usual sense, with the IR cut - off. In both
cases
 it is necessary to add to these expressions the  $ \delta$-terms, containing $
\delta$
 functions of certain external or transferred momenta. These terms in the
coordinate
 space contain the analytic dependence on certain distances. Precise
 form of the $  \delta $  - terms requires for its definition the matching with
the stirring
 force - the task not accomplished in this paper.
 \section {The Flux States }
In the above considerations we accounted for viscosity in a somewhat symbolic
 sense.
 Namely, we have used the inviscid Hopf equations, but assumed that at the
 very high momenta our power-like correlation functions start to drop rapidly
 because of the viscosity, which was assumed as the origin of the ultra- violet
cut-off.\par
 It is clear that the precise nature of this cut-off must be important. For
 instance, we can imagine an "elastic" cut-off  which preserves the hamiltonian
structure
 and the "inelastic" one which introduces dissipation. It is clear that they
 should correspond to different physics.\par
 In this section we' ll try to impose the dissipation condition on the theory.
 The basic idea is roughly the same as in the classical Kolmogorov treatment of
 turbulence.
 Namely, we will assume (self-consistently) that integrals of motion are
dissipated
 at the UV cut-off, and in the inertial range of momenta they are just
 transferred.
 It implies that the flux in the momentum space of conserved integrals must be
 constant.
 One can say that while Gibbs distributions are uniform on the surfaces of
fixed
 values of conserved quantities, the turbulent distributions are located on the
 surfaces of the constant fluxes of the corresponding quantities.\par
 The constant flux in the momentum space is also a familiar object in the other
 part of physics. It is responsible for the anomalies in the quantum field
theory.
 When one considers, for example, massless fermions in the electromagnetic
 field the chirality (the number of left minus the number of right Dirac
particles),
  is not conserved due to axial anomaly. This happens because the ultra-violet
 regularization breaks conservation of the axial current.
 After the chirality is injected at the UV cut-off, it propagates in the
momentum
 space to the physical region.
 There is, therefore, a clear and useful analogy with what we are
discussing.\par
 It has been noticed by Kraichnan [6], that in two dimensions the most
important
 flux is that of enstrophy (we will comment on that later).
 Let us derive the constraint on the theory which follows from the enstrophy
 conservation. Consider first the enstrophy contained in the modes with
definite
 momentum:
 \begin{equation}\label{}
h \left( \vec{k} \right) ~=~<\omega \left( \vec{k } \right)     ~\omega\left(
\vec{-k} \right)>
\end{equation}
 Without viscosity and external force,   $ h  \left( \vec{k} \right) $would
satisfy the continuity equation
 in the momentum space. When these factors are accounted for, the equation
  becomes:
 \begin{equation}\label{}
\dot{h}\left( k \right)+{\partial \over \partial k_{i}}{J^{(h)}_{i}}=\nu
k^{2}h\left( k \right)+\Phi\left( k \right)
\end{equation}
 Here $ \Phi\left( \vec{k} \right)$ is a contribution of the external forces
which is non-zero only for $ |\vec{k}|\sim {1 \over L}$.
 Let us notice also, that in isotropic turbulence the only non-zero component
of $ \vec{J}^{\left( h \right)}$
 is the radial one $ J^{\left( h \right)}\left( k \right)$.
 Let us now chose the momentum q  lying in the inertial range and integrate
 over $ |\vec{k}|~>~|\vec{q}|$.
 Since in the steady state $ \dot{h}\left( k \right)~=~0$
 we get
  \begin{equation}\label{}
-J^{(h)}\left( q \right)=~\nu\int_{|k|>|q|}{k^{2}h\left( k \right)d^{2}k}
\end{equation}
 Now comes an important point - suppose that the RHS of (51) is UV- divergent
 and defined by $ |k|\sim {1 \over a}$.
 Then the  q -dependence of RHS can be neglected.
 As a result, we obtain:
 \begin{equation}\label{}
J^{(h)}\left( q \right)\approx  - \nu \int{k^{2}h\left( k
\right)}d^{2}k=\mbox{const}
\end{equation}
  for
 $ {1 \over L}\ll|q|\ll{1 \over a}$.
 The meaning of this constant flux condition is transparent. It just says that
 enstrophy is dissipated only at $ |q| \sim {1 \over a}$
 and hence, the conservation law implies that its flux must be constant in the
 inertial range. This is one of matching conditions we must impose on the
inviscid
 solutions.
If we express $ \dot{\omega\left( \vec{q} \right)}$ by means of eq. (35) we
get:
 \begin{equation}\label{}
J^{(h)}\left( q \right)=~-\int_{|k|>q}{<\dot{\omega \left( k
\right)}\omega\left( -k \right)>}=~\int_{|k|<q}{<\dot{\omega\left( k
\right)}\omega\left( -k \right)>}
\end{equation}
 (we used here the fact, that (35) conserves total enstrophy).
 We see from here that all  $ \vec{k}- s$ from the inertial range give zero
contribution.
 At the same time, using our general conjecture concerning the infrared terms
 (46) we obtain:
  \begin{equation}\label{}
J^{(h)}\left( q \right)\propto {1 \over
L^{\Delta_{\dot{\omega}}+\Delta_{\omega}}}
\end{equation}
  The right hand side of (51) is independent of L and thus we obtain the
matching
 condition:
 \begin{equation}\label{}
\Delta_{\omega}~+~\Delta_{\dot{\omega}}~=~0
\end{equation}
 If we had used the first term in ( 52), where  $ k \sim {1 \over L}$ are
absent, the IR contribution
 would come from the integral (35) for  $ \dot{\omega}$ , yielding the same
result.
 If we recall, that:
 $ \Delta_{\omega}~=~\Delta_{\psi}~+~1$ and ~$
\Delta_{\dot{\omega}}~=~\Delta_{\phi}+2$
 we get a following condition for the enstrophy flux state:
 \begin{eqnarray}
[\psi]~[\psi]~=~[\phi]+\ldots \nonumber\\ ~\Delta_{\psi}+\Delta_{\phi}+3=0
\end{eqnarray}
 This result heavily depends on our conjectures about IR - contributions,
namely  on  the fact that  $ <\omega\dot{\omega}>$ infrared part is non-zero.
Had we considered the energy flux state, the formula ( 55) would be replaced
by:
 \begin{equation}\label{}
\Delta_{\psi}~+~\Delta_{\phi}~+~2~=~0
\end{equation}
  Apart from the energy and the enstrophy  there are also higher conserved
integrals of the type:
 \begin{equation}\label{}
I_{n}~=~\int{\omega^{n}\left( x \right)~d^{2}x}
\end{equation}
 Before commenting on these integrals, let us explain why the formulas (56) and
(57) do not contradict each other. (After all, energy and enstrophy are both
conserved in the inviscid system). The apparent contradiction is removed by the
fact that in the enstrophy flux state the integral for the energy dissipation,
analogous to (50) is given by:
 \begin{equation}\label{}
J^{(\varepsilon)}\left( q \right)=~\nu \int_{|k|>q}{d^{2}k<\omega\left( k
\right)\omega\left( -k \right)>}
\end{equation}
 Contrary to (50), (58) is not UV - divergent and the dissipation is scale -
dependent in the inertial range.\par
 In this case we can't use (51) anymore.
No obvious matching condition arise in this case, since (58) tells us that as $
\nu\rightarrow 0 $ the energy flux tends to zero, while the enstrophy flux
persists. This is essentially the picture,
 advocated by Kraichnan [6].

As we turn to the integrals $I_{n}$ we have to decide, whether they have
 nonzero flux and whether this gives us any new information. We are unable at
this point to give complete analyses of this question and will present some
non-rigorous estimates, leaving complete resolution for the future work.
We have:
 \begin{equation}\label{}
\dot{I}_{n}~=~\nu \int{<\nabla ^{2}\omega\left( x \right)\omega^{n-1}
\left(x
\right)>d^{2}x}
\end{equation}
 The main problem is how to regularize this expression. We must use OPE in
order to define $ \omega^{n-1}$ .If n   is even,  which is imposed by parity
conservation, we get
 \begin{equation}\label{}
\omega^{n-1}\left( x \right)\infty~\Psi\left( x \right)~+~\cdots
\end{equation}
 and as a result:
 \begin{equation}\label{}
\dot{I_{n}}\propto \int{<~\triangle \omega~~\psi>}\propto J^{(\varepsilon)}
\end{equation}
We picked up  $ \psi$ operator only in(61) since due to orthogonality of
primary operators, all others will not contribute.
  This implies that the   $ I_{n}$- transfer is zero in the inviscid limit and
there are no extra constraints on the theory coming from it (apart, may be from
some inequalities). However, this line of arguments must be considered in more
details (with careful point - splitted definition of all parties involved,
before the last conclusion could be trusted). In particular,the other groupings
of $ \omega$ -s may involve some other operators. Much work here is to be
done.\par An interesting related question is that of the inverse cascades of
higher integrals. \footnote{I acknowledge useful discussion of this point with
G. Falcovich and A. Hanany} Strong infrared divergency in their dissipation
makes them very similar to the energy and opens this  possibility
for $ I_{n}$ . Whether this  is the case is an  unsolved  question.
Unfortunately, any kind of the naive pseudophysical arguments is misleading
 here.\par
To conclude this section, let us discuss the physical picture of the flux
state. It is not exactly the Kolmogorov's one.
We have constant flux of conserved quantity through the inertial range. In the
Kolmogorov's mechanism the transfer occurs by the interaction of three modes
with all wave vectors lying in the inertial range. Such transfer is absent in
our picture. Instead we have the triad interaction in which one of the modes is
infrared. We can say that the infrared (large scale) modes serve as catalyzers
for the flux. This is an important qualitative prediction of our theory, which
can be tested.
 \section{Possible spectra of conformal turbulence.}
 Let us discuss solutions of conformal turbulence with constant enstrophy flux.
The analyses of the previous sections indicates that we must start from the
CFT, satisfying certain constrains which follow from the Hopf equation and
constant flux condition. The first constraint tells us that if we have the
fusion rule:
 \begin{equation}\label{}
 [\psi]~[\psi]~=~[\phi]~+~\cdots
\end{equation}
 where $ \phi$ is a minimal dimension operator in the product (62), then it
must be either $ \Delta_{\phi}>2\Delta_{\psi}$ or
   $ \phi$ - operator must be degenerate at the level two.\par
The second constraint depends on the assumption about vacuum expectation
values. The simplest version of the theory, when these values are zero gives:
 \begin{equation}\label{}
\Delta_{\phi}~+~ \Delta_{\psi}~=~-3
\end{equation}
 More options arise if we assume that if $ <O_{n}>~\neq~0$
 In this case the correlation functions $ \omega\omega$
and $ \dot{\omega}\omega$
are determined not by the  unit operator  but by the minimal dimension
operators. As a result, if we look at OPE, we find:
 \begin{eqnarray}
<\omega\left( z \right)\omega\left( 0 \right)>\propto \left( \bar{z}z
\right)^{\Delta_{\phi}-2\Delta_{\omega}} ~<\phi> \nonumber\\
\Delta_{\omega}=\Delta_{\psi}+1
\end{eqnarray}
 and
 \begin{equation}\label{}
<\dot{\omega}~\omega>\propto\left( \bar{z}z \right)^{-\left(
\Delta_{\omega}+\Delta_{\dot{\omega}} \right)}\left( \bar{z}z
\right)^{\Delta_{\chi}}
\end{equation}
 where $ \chi$ is a minimal dimension operator in the product
 $ [\psi]~[\psi]~[\psi] .$
 So, in this case the constant flux condition is replaced by:
 \begin{equation}\label{}
\Delta_{\phi}~+~\Delta_{\psi}~-~\Delta_{\chi}~=~-3
\end{equation}
 while the energy spectrum is given by:
 \begin{equation}\label{}
E\left( k \right)~ \sim~k^{4\Delta_{\psi}-2\Delta_{\phi}+1} \end{equation}
 One can also consider intermediate possibilities when some of the operators
have vacuum averages and some do not.\par
At present we do not have clear idea how the choice should be made. It may
depend on the external conditions and the way turbulence is generated.\par
Below we will examine a simplest possible model, the "minimal" minimal model.
By that we mean the theory which satisfies our requirements and has the
smallest number of primaries. If we associate each primary operator with the
certain type of motion then it is indeed the simplest (but by no means the
only) type of turbulence.
Let us look at the minimal models of the type (2,2N +1), and assume, that
vacuum expectation values are absent.
In this models we have the set of N primary operators $ [\psi_{s}]$,
with $s=1, ~\cdots N$.
 The fusion rules are:
  \begin{eqnarray}\label{}
[\psi_{s_{1}}]~[\psi_{s_{2}}]~=~[\psi_{s_{3}}] +[\psi_{s_{3}-2}]+ \ldots
\nonumber\\ s_{3}=min\left( s_{1}+s_{2}-1,2N+1-\left( s_{1}+s_{2}-1 \right)
\right)
\end{eqnarray}
 The dimensions are given by:
 \begin{equation}\label{}
\Delta_{s}~=~-{\left( 2N - s \right)\left( s-1 \right) \over 2\left( 2N +1
\right)}
\end{equation}
If we identify the stream function $  \psi$ with the primary field  $
\psi_{S}$ with some S , then the flux condition gives:
 \begin{equation}\label{}
\Delta_{s}+\Delta_{2s-1}~=~-3
\end{equation}
 (if $ 2\left( 2s - 1 \right)~<~2N+1). $
 This diophantine equation has unique solution, s=4; N=10 .
Thus, we conclude that the minimal turbulence is described by the (2~,21)
minimal model.
Anomalous dimensions in this case are
 \begin{eqnarray}\label{}
\Delta_{\psi}~=~\Delta_{4}~=~-{8 \over 7}\\
\Delta_{\phi}~=~\Delta_{7}~=~-{13 \over7}
\end{eqnarray}
 As a result we obtain in this case the energy spectrum
 \begin{equation}\label{}
E\left( k \right)~\infty~k^{-{25 \over7}}
\end{equation}
 It seems to be consistent with observations. [7]\par
An interesting question if this model is parity conservation. There is no a
priori reason why we should exclude parity breaking solutions from the
consideration.
Let us look what is the situation in our model.
The stream function  $ \psi$ is a pseudoscalar. That means that if parity is
conserved, the 3 - point function
must contain vector products. But in CFT this is impossible: all 3-point
functions are simple powers of distances.\footnote{Some time ago A. Migdal
conjectured that in 3d turbulence T- invariance forbids the 3-point functions.
Whether it is true is an interesting open question. In our case the only thing
which can be proved is the  combined PT-invariance .}
Hence, in conformal turbulence the necessary condition for parity conservation
is:
 \begin{equation}\label{}
<\psi~\psi~\psi>~=0
\end{equation}
 This is indeed the case, since:
 \begin{equation}\label{}
[\psi_4][\psi_4]=[\psi_7]+[\psi_5]+[\psi_3]+[\psi_1]
\end{equation}
 and there is no $ [\psi_4]$  in the RHS. As we look at higher correlations the
situation is less clear. Indeed, the 5 - point functions of $  \psi$ are still
zero (by the similar arguments) while the 7 -point functions are non-zero. If
parity is conserved, they must contain vector products. It is not  obvious,
whether this can be done, or, in other words, whether we can combine conformal
blocks in the 7 - point function in a antisymmetric (under reflections) way.
This we leave for the future analyses, noticing only, that if we consider the
standard symmetric combinations of the conformal blocks, we get a very peculiar
picture in this model. namely, the parity violation, if present at all, is well
hidden - one must go to the seven - point functions in order to notice it!\par
Let us point out that parity is automaticaly conserved in the non-minimal
models. In this case  there exists infinite number of  the degenerate
operators, forming closed algebra preserving parity [4].

 \section{Conclusion.}
There are infinitely many other solutions of the flux conditions, the fixed
points with larger number of structures. They have been considered in [8]. It
is hard to say, whether all matching conditions have been exploited and whether
we indeed have infinite number of discrete fixed points. If so, the turbulent
fluid has infinite discrete regimes and must be able to jump from one regime to
the other, generating new quasistable structures. we are not in a position yet
to prove this conjecture, but it is certainly within the scope of our
methods.\par
As far as observations are concerned, the best way to check conformal
invariance is to study momentum representation of the correlation functions
when all the momenta and all their partial sums belong to the  inertial range.
In this way we avoid the infrared region which is scarcely understood.\par The
situation is analogous to that in quantum chromodynamics - there we know
dynamics at high momenta and quite ignorant in the infrared. Nevertheless, QCD
has been tested.\par
The theory of conformal turbulence gives explicit and unambiguous predictions
concerning these type of correlators. They are expressed in terms of
hypergeometric functions, and, more importantly have specific "fusion rule"
structure when one momentum is getting much larger than the others. The most
direct way to test the theory is to check this fusion property. For example,
one takes the 4 - point function and reduce it to the 3 - point function by
taking one large momentum. This 3 - point function in the momentum space is a
simple hypergeometric function, which becomes just a product of two powers if
one does the fusion procedure again.
This test is especially important, since we are not certain which solution of
conformal turbulence is realized in a given experiment.
Since the structure of
fusion rules define the theory the test is capable to answer this question.
On the other hand, may be the most interesting part of the theory is the
infrared one, still hidden from us. With further work it might become possible
to get equations, governing the behavior of the vacuum expectation values of
  $<O_{n}>$ in space and time. This is the problem analogous to the one of
equation of state in critical phenomena.
In any case, the theory presented here is but a first step in the long future
investigations.
 \section{Acknowledgements}
 I am grateful to A.Migdal, E.Siggia, V.Yakhot and A.Zamolodchikov for many
important remarks on this work. I also thank D. Makogonenko for her help in
preparation of this article.
 This work was partially supported by the National Science Foundation under
contract PHYS-90-21984.
 \eop
\appendix{REFERENCES\par [1] A.Polyakov ~Preprint ~PUPT-1341 ~(1992)
(Princeton)\par[2]~A.Monin ~A.Yaglom,~"Statistical ~Fluid ~Mechanics"~MIT
Press,~Cambridge (1975)\par [3] J.Kogut, K.Wilson ~Phys. Rep., 12C, 2,
(1974)\par [4]~A. Belavin, A. Polyakov, A. Zamolodchicov,~Nucl. Phys. B241, 33,
(1984)\par [5] Al. Zamolodchicov, Nucl. Phys. B358, 497, (1991)\par [6] R.
Kraichnan, Phys. of Fluids 10, 1417, (1967)\par [7] B. Legras, P. Santangelo,
R. Benzi, Europhys. Lett. 5 (1) 37, (1988)\par[8] D. Lowe Preprint PUPT-1362
{}~(1992) (Princeton)\par G. Falcovich, A. Hanany, Weizman Institute Preprint
(1992)\par V. Pokrovsky, private communication}
\end{document}